\documentclass[twoside, transmag]{IEEEtran}

\usepackage[english]{babel}
\usepackage{siunitx}
\DeclareSIUnit\torr{torr}
\DeclareSIUnit\ppm{ppm}

\usepackage{amsmath} 
\usepackage[nameinlink,compress]{cleveref}
\crefname{figure}{Fig.}{Figs.} %To be consistent with IEEE that wants "Fig. 1".
\usepackage[noadjust, nospace]{cite} %no space before citations
\usepackage{booktabs}
\usepackage{array} %to increase fonts in tables with arraystretch 
\usepackage{stackengine} %to increase single row height in tables 
\usepackage{graphicx}
\usepackage{braket}
\usepackage{url}

\usepackage[]{subcaption} %both packages enable subfigures

\usepackage{float}	%more positioning options (es: H)

\begin{document}
	
%\bstctlcite{MyBSTcontrol}		%Modify IEEEtran to avoid URLs (together with additional custom_IEEEtran.bib file)

%\newcommand{\titletext}{Loaded microwave cavity for compact vapor-cell clocks}

\title{Loaded microwave cavity for compact vapor-cell clocks} %Title of paper

\author{Michele Gozzelino, Salvatore Micalizio, Claudio E. Calosso,  Aldo Godone, Haixiao Lin and Filippo Levi 
	\IEEEcompsocitemizethanks{\IEEEcompsocthanksitem M. Gozzelino, S. Micalizio, C. E. Calosso,  A. Godone, H. Lin, and F. Levi are with Istituto Nazionale di Ricerca Metrologica, INRIM, Torino, Italy.\protect\\
		E-mail: s.micalizio@inrim.it
		\IEEEcompsocthanksitem H. Lin is with Key Laboratory of Quantum Optics, Shanghai Institute of Optics and Fine Mechanics, Chinese Academy of Sciences, Shanghai 201800 and with University of Chinese Academy of Sciences, Beijing 100049, China. Lin H. acknowledges the China Scholarship Council (CSC) and the National Natural Science Foundation of China (NSFC) under Grant No. 91536220. }}% <-this % stops a space

%\markboth{Journal of \LaTeX\,~Vol.~?, No.~?, \date{\today}}%
%{Gozzelino \MakeLowercase{\textit{et al.}}: \titletext}

\maketitle 

\begin{abstract}
Vapor-cell devices based on microwave interrogation provide a stable frequency reference with a compact and robust setup.  Further miniaturization must focus on optimizing the physics package, containing the microwave cavity and atomic reservoir.  In this paper we present a compact cavity-cell assembly based on a dielectric-loaded cylindrical resonator. The structure accommodates a clock cell with \boldmath$0.9 \, \mathrm{cm^3}$ inner volume and has an outer volume of only $35 \, \mathrm{cm^3}$. The proposed design aims at strongly reducing the core of the atomic clock, maintaining at the same time high-performing short-term stability ($\sigma_y(\tau) \leq 5\times 10^{-13} \,\tau^{-1/2}$ standard Allan deviation). The proposed structure is characterized in terms of magnetic field uniformity and atom-field coupling with the aid of finite-elements calculations. The thermal sensitivity is also analyzed and experimentally characterized. We present preliminary spectroscopy results by integrating the compact cavity within a rubidium clock setup based on the pulsed optically pumping technique. The obtained clock signals are compatible with the targeted performances. The loaded-cavity approach is thus a viable design option for miniaturized microwave clocks. 

\end{abstract}

\section{Introduction}
\label{sec:intro}
Vapor-cell atomic clocks provide excellent frequency stability together with low volume and power consumption. They are employed both on ground and in space applications~\cite{Riley2019}. Laser-based devices, working either with continuous or pulsed schemes, have shown state-of-the-art stability levels in their class, making them attractive options for next generation global navigation systems~\cite{Godone2015,Camparo2015,schmittberger2020}. Several industries have shown increasing interest for the pulsed optically pumped (POP) Rb clock, in virtue of its demonstrated performances and mature technology~\cite{Arpesi2019}. 

For industrial and spaceborne applications, a compact and light physics package is of high importance, not only for weight and volume considerations, but also for the possibility to reduce the overall power consumption and to improve the mechanical design. Moreover, for space applications, the mid and long-term stability is of greater relevance than having state-of-the-art short term performances ($\simeq$\num{1.5e-13}$\,\tau^{-1/2}$)~\cite{Micalizio2012a,Bandi2014}. The development of a  compact physics package goes in this direction, providing easier temperature stabilization and mitigation of temperature gradients, compared to a distributed object. This results in foreseen improved mid-term frequency stability performances~\cite{Hudson2018}. 

This work presents a miniaturized microwave cavity suitable for POP~\cite{Micalizio2012} or continuous-wave~\cite{Bandi2014} Rb clocks, looking for a trade-off between physical dimensions and desired short-term stability performances. 

The physics package of Rubidium standards typically presents a layered structure, including several layers of magnetic and thermal shields. To reach an overall reduction the package size, the most efficient strategy is to reduce the volume of the physical core of the clock, i.e. the microwave cavity. Many high-performing clocks make use of a cylindrical cavity resonant on the $\mathrm{TE_{011}}$ mode, because of its favorable $\mathbf{H}$-field distribution, which is uniform and parallel to the cavity axis in the central region of the cavity. This features guarantees to excite the clock transition in an efficient way, increasing the contrast of the clock signal~\cite{Godone2011}. The cavity inner dimensions are designed in order to tune the resonance frequency to the atomic clock transition (\SI{6.8}{\giga\hertz} for $\mathrm{^{87}Rb}$).

One way to shrink the cavity dimensions, keeping the magnetic field resonant with the atomic clock transition, is the use of a loop-gap resonator (also called ``magnetron'' or ``split-ring'' resonator)~\cite{Froncisz1982,Sphicopoulos1987}. This approach has been proven effective both for the continuous-wave and the POP rubidium frequency standards~\cite{Stefanucci2012,Kang2015,Hao2019}. Indeed, it provided reduction of the overall cavity-cell volume up to a factor 4.5, retaining comparable cell size and clock performances, compared to the traditional cylindrical cavity.

Inserting a dielectric material inside the cavity volume can also lead to remarkable volume reduction~\cite{Howe1983,Williams1983}. Dielectric loading can also be exploited, to some extent, to increase the field uniformity in the active volume~\cite{Mett2001,Wang2019}. 

We propose a novel design solution for the cavity-cell assembly based on an alumina-loaded microwave cavity, demonstrating an external volume of only \SI{35}{\centi\meter\cubed}. %The inner loaded-cavity volume showed a reduction of a factor 10 compared to the previous prototype based on a traditional cylindrical cavity~\cite{Micalizio2012}. 
The loaded cavity still works on a $\mathrm{TE_{011}}$-like mode, ensuring favorable magnetic field uniformity and directionality. Such strong size reduction is possible by scaling also the clock cell, whose inner volume is reduced by a factor 8. As demonstrated in more detail in \Cref{sec:design} and \Cref{sec:exp}, this volume reduction partially affects the clock short-term stability, but still provides interesting performances. 

The proposed alumina-loaded cavity offers new design alternatives and thus facilitates the use of the POP technology for in-field applications. Advantages of this approach is the notable size reduction and mechanical stability, as the dielectric can serve also as self-centering spacer for the clock cell. 

The paper is organized as follows: in \Cref{sec:design} the proposed design is introduced. In \Cref{sec:FEM} the loaded-cavity main features are analyzed with the aid of Finite Element Method (FEM) analysis. Finally, in \Cref{sec:exp} the cavity is experimentally characterized, and foreseen short-term stability performances for such a cavity in a POP clock experiment are discussed. 

\section{Cylindrical loaded-cavity design}
\label{sec:design}
Reducing the cell dimensions has two effects: first, it reduces the available atomic sample volume, thus decreasing atom number and the signal-to-noise ratio. Second, the atomic population relaxation rates are increased, due to the collisions with the cell walls. The first point is not critical for many laser-based vapor-cell clocks, since they are not limited by shot-noise, rather from laser intensity and frequency noises~\cite{Calosso2020}. The second issue can be mitigated by finding an optimal buffer gas pressure for which the total relaxation rate is minimized. Indeed, since for typical operational temperatures the main transverse relaxation-rate contribution is due to spin-exchange collision between Rb atoms, there is margin to increase the pressure of the buffer gas without significantly enhancing the total relaxation rate~\cite{Micalizio2009}. 

Given the last considerations, for a traditional buffer-gas mixture composed of Ar and $\mathrm{N_2}$ (1.6:1 ratio)~\cite{Vanier1982}, the total buffer-gas pressure is set to \SI{40}{\torr} (\SI{53.2}{\hecto\pascal}). At this pressure, and for a typical operational temperature of \SI{62}{\degreeCelsius}, the transverse relaxation rate $\gamma_2$ is minimized ($\gamma_2\simeq\SI{400}{\per\second}$)~\cite{Vanier1989}. Compared to the case of a \SI[product-units = repeat]{2 x 2}{\centi\metre} cylindrical cell with \SI{25}{torr} of buffer gas ($\gamma_2 = \SI{280}{\per\second}$), this corresponds to a 1.4 factor increase in the transverse relaxation rate. The $\gamma_2$ rate causes a decay of the atomic signal, ultimately limiting the interrogation time length ($T\approx\gamma_2^{-1}$), so we expect the optimal Ramsey time to be reduced roughly by the same factor ($T\simeq\SI{2.4}{\milli\second}$ versus $T\simeq\SI{3.4}{\milli\second}$). Finally, a reduction of the interrogation time lowers the atomic line quality factor, impacting the short-term stability. We consider these values a reasonable trade-off between expected performance and physics package size reduction. 
\begin{figure}
	\includegraphics[width=\columnwidth]{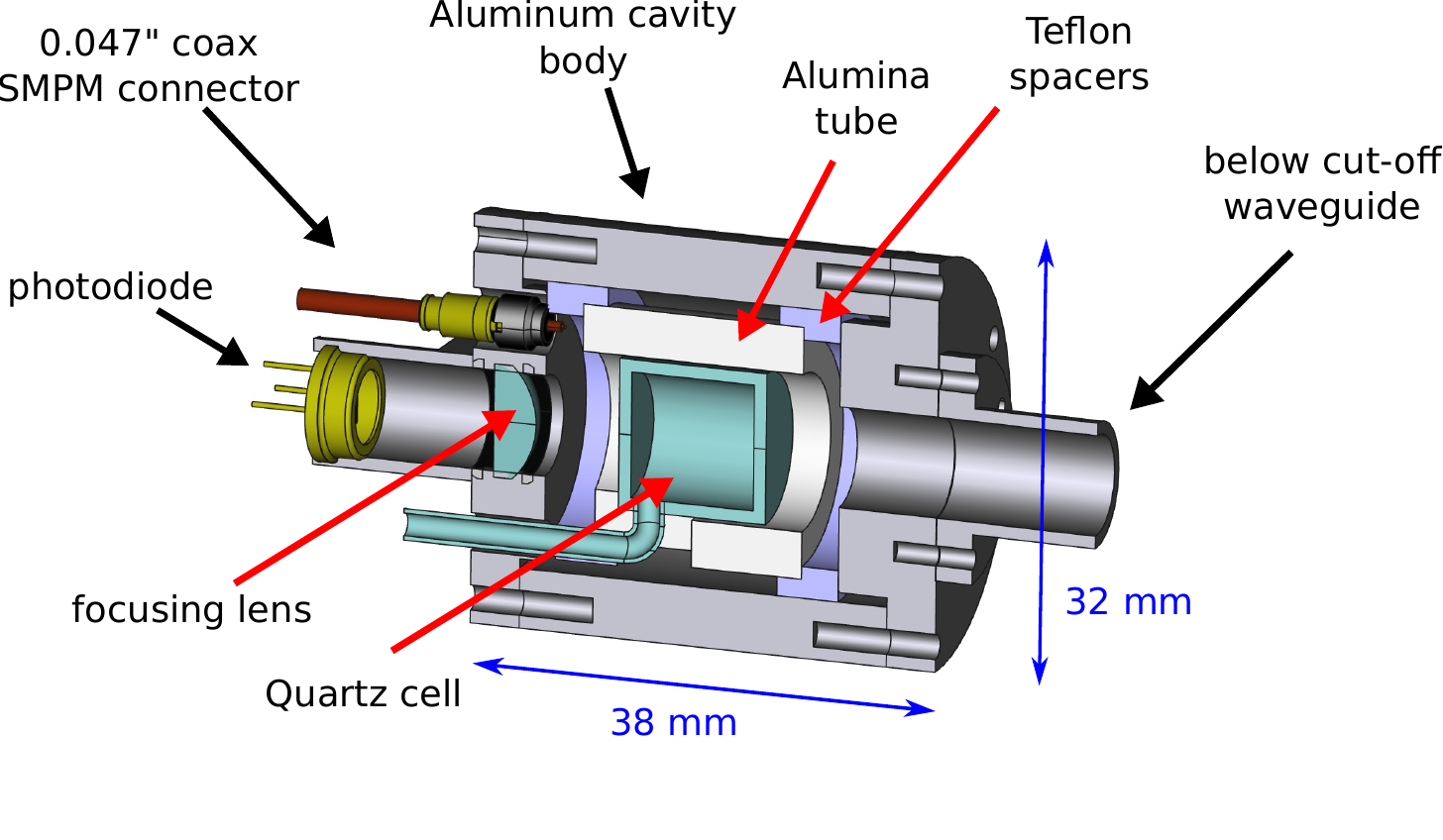}
	\caption{\label{fig:CAD} Rendering of the cavity-cell assembly (cross-section view).}
\end{figure}

The complete setup is shown in \Cref{fig:CAD}. The dielectric has a cylindrical shape, to preserve as much as possible the symmetry of the system. A small indent is introduced to allow the cell stem to exit the cavity volume (a cold point is desirable to induce metallic rubidium condensation outside the cavity volume). The alumina tube is centered with the aid of two Teflon rings that increase the mechanical tolerance. The cavity is made of aluminum and consist of a cylindrical body and two endcaps fixed by a set of screws. The end-caps of the cavity present two circular apertures, with \SI{9}{\milli\meter} diameter, for optical access. These apertures are followed by cylindrical wave-guides with the same diameter. With such a geometry, the $\mathrm{TE_{011}}$ mode frequency is below the waveguide cut-off frequency and power attenuation of at least 40 dB for the evanescent microwave field is obtained with a total waveguide length of \SI{12}{\milli\meter}, making the clock frequency instability due to microwave leakage negligible. Also in this case, the reduction of the cell size and, consequently, of the needed optical access allowed us to scale the waveguides volume, maintaining the same attenuation of the unloaded design. On one side of the cavity, a magnetic coupling loop (created with a short-circuited SMPM coaxial cable) provides the microwave excitation. The system is completed with a plano-convex lens which focuses the laser light onto a \SI[product-units=repeat]{2.6x2.6}{\milli\meter} Si photodiode. 

\subsection{Finite element analysis (FEA)}
\label{sec:FEM}
The resonance frequency and spatial distribution of the cavity modes of interest are analyzed with the aid of the finite-element-method software tool CST-studio~\cite{CSTstudio}. %Since, regardless of the loading, the cavity transverse electric (TE) and transverse magnetic (TM) modes clearly resemble the unperturbed ones, we will use the same notation throughout the paper.

At first, the geometry of the cavity, cell and loading material are determined, so that the $\mathrm{TE_{011}}$-mode frequency matches the atomic frequency. A check of the nearest resonance modes is made, to verify that they do not overlap with the main cavity mode. 

\begin{table}
\caption{\label{tab:modes} Resonance frequencies and $Q$-factor for the main cavity modes (without clock cell).}
\centering %centra la tabella
\renewcommand{\arraystretch}{1.5}
\resizebox{0.49\textwidth}{!}{
	\begin{tabular}[]{l c c c c} 
		\toprule
		mode  & $\nu_0$ (FEA)\footnote{Assuming conducibility of Al $\sigma = \SI{3.8e7}{\siemens\per\meter}$ } & $Q_i$ (FEA) & $\nu_0$ (exp.) & $Q_i$ (exp.) \\ 				%heading
		\hline  
		%\addlinespace[1ex]  %add some space as superscripts touch the hline
		$\mathrm{TE_{011}}$  	 & \SI{7.17}{\giga\hertz}    &	 5200  &  \SI{7.00\pm0.01}{\giga\hertz}	& $4000\pm200$ \\	
		
		$\mathrm{TM_{111}}$  	 & \SI{7.58}{\giga\hertz}    &	 5100  &  \SI{7.50\pm0.01}{\giga\hertz}	& $3200\pm200$ \\	
		
		$\mathrm{TM_{011}}$  	 & \SI{6.20}{\giga\hertz}    &	 4200  &  --	& --	\\ 
		\bottomrule
	\end{tabular}}
\end{table}

From the FEA analysis, given their spatial distribution, the two nearest modes are recognized as the $\mathrm{TM_{011}}$ and $\mathrm{TM_{111}}$. They lie at least \SI{400}{\mega\hertz} away from the eigenmode of interest for the clock operation. Conveniently, the dielectric loading completely lifts the degeneracy of the $\mathrm{TM_{111}}$ and $\mathrm{TE_{011}}$ modes, thus no mode-choke for the $\mathrm{TM_{111}}$ mode is necessary. In \Cref{tab:modes} a list of the simulated resonance frequency modes and relative intrinsic $Q$-factors (considering also the dielectric losses from the loading materials) is shown for the loaded microwave cavity without the clock cell. The simulated values are compared to the measured one. As it will also be noted in \Cref{sec:exp}, the discrepancy in the absolute frequency is mainly due to the value of the dielectric constant of the alumina at \SI{6.8}{\giga\hertz}, which resulted \SI{3}{\percent} higher than the value given from the manufacturer~\cite{morgan-datasheet}.

Second, the uniformity and directionality grades of the $\mathrm{TE_{011}}$ mode is evaluated. In particular a ``uniformity coefficient'', introduced in \cite{Godone2011}, is used to characterize the amplitude variations of the $H_z$ component over the active volume ($V_a$). To avoid ambiguity,	the latter is taken coincident with the cell inner volume. Compared to \cite{Godone2011}, the definition has been slightly modified to allow the maximum of the magnetic field to lie away from the center of the cavity (due to asymmetries in the cell and loading materials):
\begin{equation}
u = \frac{\frac{1}{V_a}\int_{V_a}^{}H_z(\mathbf{r})^2 \; dV }{\displaystyle\max_{\mathbf{r}\in V_a} \lbrace H_z(\mathbf{r})^2\rbrace}
\end{equation} 
where $\mathbf{r}$ is the spatial coordinates vector. In this way \mbox{$0<u<1$}, with $u=1$ corresponding to a perfectly uniform field distribution. 
Following \cite{Stefanucci2012}, we also report the ``orientation coefficient'' $\xi$, defined as:
\begin{equation}
\xi = \frac{\int_{V_a}^{}H_z(\mathbf{r})^2 \; dV }{\int_{V_a}^{} |\mathbf{H}(\mathbf{r})|^2 \; dV }
\end{equation}
This coefficient is a figure of merit of the orientation of the $\mathbf{H}$ field along the quantization axis $z$. A high orientation factor ($\xi$ close to 1), minimizes the excitation of the $\sigma$-transitions ($\Delta F=1$, $\Delta m_F=1$), that can cause unwanted cavity pulling on the clock transition~\cite{Gerginov2014}. For completeness, we also report the filling factor, which is mostly important for active oscillators (or POP with microwave detection~\cite{Godone2004}), expressing the degree of coupling between the microwave field and the atomic sample:
\begin{equation}
\eta'= \frac{\left( \int_{V_a}^{}H_{z}(\mathbf{r}) \; dV \right)^2}{V_a\int_{V_c}^{}|\mathbf{H}(\mathbf{r})|^2 \;dV}
\end{equation}
where $V_c$ is the inner cavity volume.
\begin{figure}[]
	\centering
	\begin{subfigure}[b]{0.3\textwidth}
		\centering
		\includegraphics[width=\textwidth]{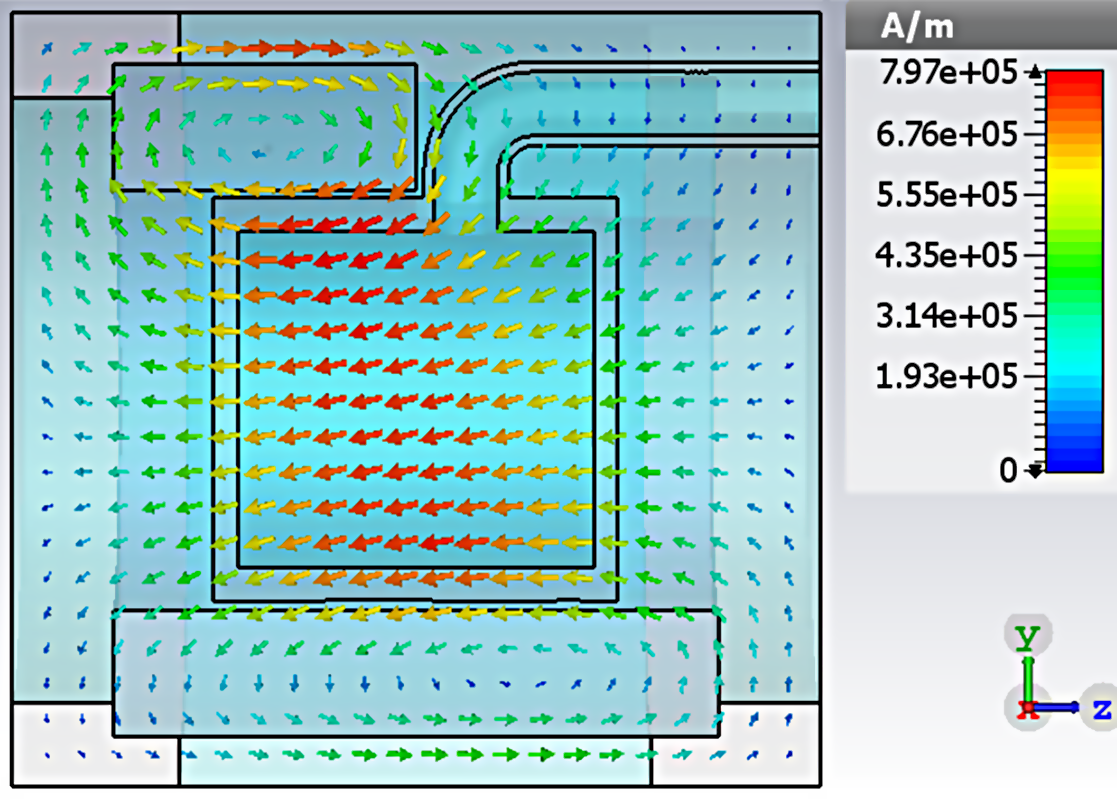}
		\caption{$\mathbf{H}$ field in the $y$-$z$ plane.}
		\label{fig:H}
	\end{subfigure}
	\hfill
	\begin{subfigure}[b]{0.3\textwidth}
		\centering
		\includegraphics[width=\textwidth]{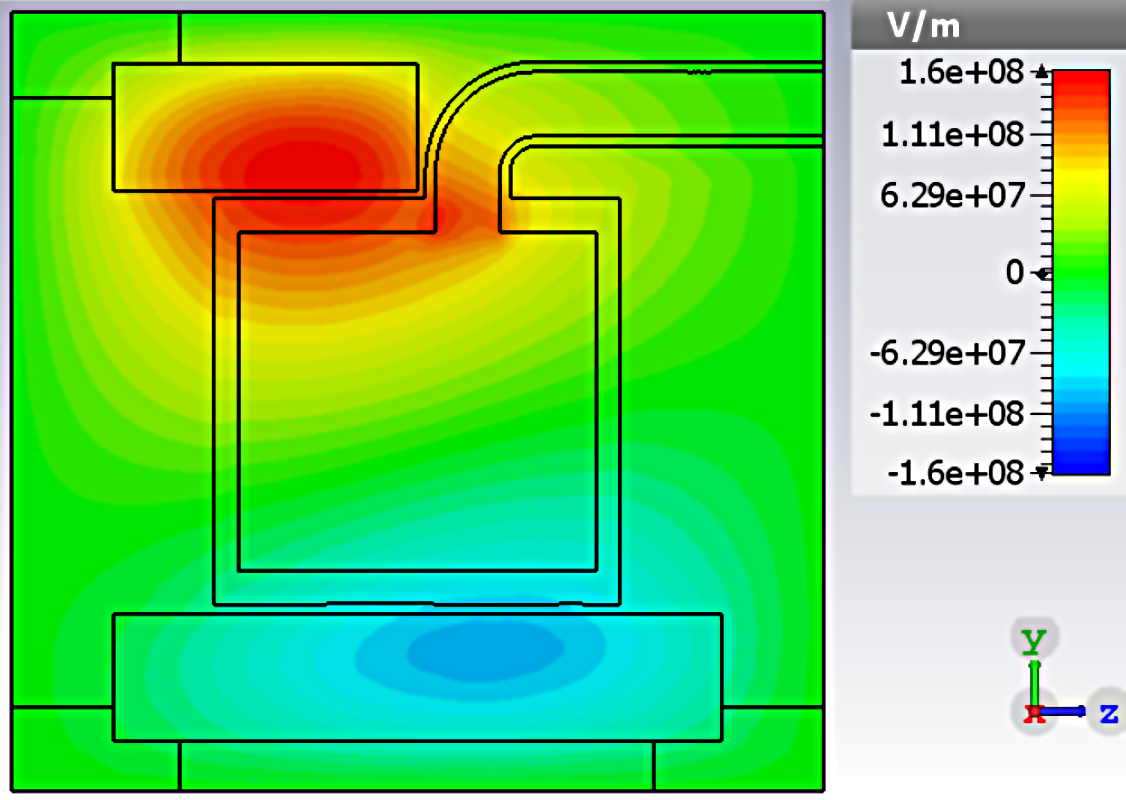}
		\caption{$E_x$ component in the $y$-$z$ plane.}
		\label{fig:E_x}
	\end{subfigure}
	
	\caption{Magnetic and electric field distribution inside the cavity volume for the $\mathrm{TE_{011}}$ mode.}
	\label{fig:FEM}
\end{figure}
In \Cref{fig:H}, the $\mathbf{H}$ field lines and amplitude for the proposed loaded-cavity assembly are shown in the \mbox{$y$-$z$} plane (the pumping/detection laser propagates along the $z$-axis). We can notice a rather good uniformity of the field component over the cell volume. We point out that along this plane we have the maximum field distortion due to the presence of the stem, while on the \mbox{$x$-$z$} plane we have better uniformity. Looking at the field lines distribution, we notice that a gap between the alumina tube and the cavity wall is not only preferable in terms of mechanical tolerances, but it also increases the mode uniformity. Indeed, due to the high dielectric constant of $\mathrm{Al_2O_3}$, the electric field is strongly concentrated inside the dielectric volume~\cite{joannopoulos2008} (see \Cref{fig:E_x}) and the magnetic field lines need a dielectric-free volume to concatenate the electric field lines. 

In \Cref{tab:coefficients} the previously introduced coefficients for the proposed configuration are reported and compared to other cavity-cell assemblies present in the literature. We can observe that the uniformity is increased compared to the case of the traditional cylindrical cavity while the orientation is comparable to other kind of compact assemblies. Finally, the filling factor is reduced compared to the case of the unloaded cavity, reducing cavity-induced sensitivities such as cavity-pulling, but still high enough to provide sufficient coupling between the field and the atomic sample to achieve efficient clock interrogation. \\

\begin{table}[t]
	\caption{\label{tab:coefficients} Comparison of different published cavity-cell assemblies in terms of magnetic field uniformity and orientation. }
	\centering %centra la tabella
	\renewcommand{\arraystretch}{1.5}
	\resizebox{0.4\textwidth}{!}{
		\begin{tabular}[]{l c c c} 
			\toprule
				cavity type	&	$u$	&	$\xi$	&	$\eta'$	 \\ 				%heading
			\hline
			%\addlinespace[1ex]
			cylindrical~\cite{Godone2011,Micalizio2012}	&	0.59	&	0.92	&	0.38	\\
			magnetron~\cite{Stefanucci2012}	&	n/a		&	0.87	&	0.14	\\	
			magnetron~\cite{Hao2019}		&	n/a		&	0.90	&	n/a		\\	
			$\mathrm{Al_2O_3}$-loaded (this work)&	0.82	&	0.76	&	0.20		\\	
			
			\bottomrule
	\end{tabular}}
\end{table}

\subsubsection*{Thermal sensitivity}
For clock applications, the stability of the cavity mode frequency is of paramount importance, as it can impact the clock frequency through cavity-pulling~\cite{Micalizio2009}. One of the main parameters of influence is temperature~\cite{Godone2011,Wang2019}, which can change the resonance frequency $\nu_c$ by thermal expansion and by affecting the dielectric properties of the materials. %It is thus interesting to evaluate the cavity eigenfrequency sensitivity to geometric and physical parameters which in turn depend on temperature. %Given the known temperature coefficients, we can then estimate the total temperature sensitivity of the cavity resonance. 

The total contribution to the sensitivity of the cavity resonance frequency to a temperature variation $\Delta T$, for small variations from the operational setpoints, can be expressed as a sum of terms:
\begin{equation}
		\frac{1}{\nu_c}\frac{\Delta \nu_c}{\Delta T}  = \sum_{k}^{} \frac{x_k}{\nu_c} \frac{\Delta \nu_c}{\Delta x_k}   \alpha_k  + \sum_{i}^{}   \frac{\epsilon_i}{\nu_c} \frac{\Delta \nu_c}{\Delta \epsilon_i}  \beta_i
\end{equation}
where $x_k$ are the geometric dimensions of the cavity and loading materials and $\alpha_k = \frac{\Delta x_k}{x_k\Delta T} $ the corresponding linear thermal expansion coefficient;  $\epsilon_i$ is the dielectric constant of the $i$-th material inside the cavity and $\beta_i = \frac{\Delta \epsilon_i}{\epsilon_i\Delta T}$ the related thermo-dielectric coefficient.  
\begin{table}[b]
	\caption{\label{tab:temp_coeffs} Major contributions to the cavity thermal sensitivity expressed in relative terms. The geometric and thermo-dielectric contributions are listed in the upper and lower part of the table respectively, in decreasing order of importance.}
	\centering %centra la tabella
	\renewcommand{\arraystretch}{1.5}
	\resizebox{0.49\textwidth}{!}{
		\begin{tabular}[]{l c c c} 
			\toprule
			\hline
			\addlinespace[2pt]  %add some space as superscripts touch the hline	
			 	$x_k$ contribution 		& \addstackgap[1pt]{ \large $\frac{x_k}{\nu_c} \frac{\Delta \nu_c}{\Delta x_k}  $ }	& $\alpha_k$ & \large $\frac{1}{\nu_c}\frac{\Delta \nu_c}{\Delta T}$ 	 \\ 	
			 		&	& $(\si[per-mode=symbol]{\ppm\per\kelvin})$	& $(\si[per-mode=symbol]{\ppm\per\kelvin})$ \\	%heading
			\hline
			%\addlinespace[1ex]
			cavity radius				&	-1.04	&  23	&	-23.9	\\
			cavity length				&	-0.13	&  23	&	-3.0	\\	
			$\mathrm{Al_2O_3}$ thickness&	-0.22	&  6.9	&	-1.5	\\	
			$\mathrm{Al_2O_3}$ length	&	-0.07	&  6.9	&	-0.5	\\
			\hline
			\addlinespace[4pt]  %add some space as superscripts touch the hline	
			\hline
			\addlinespace[2pt]
				$\epsilon_i$ contribution 		& \addstackgap[1pt]{ \large $\frac{\epsilon_i}{\nu_c} \frac{\Delta \nu_c}{\Delta \epsilon_i} $ }	& $\beta_i$ &  \\ 				%heading
				&	& $(\si[per-mode=symbol]{\ppm\per\kelvin})$	&  \\	%heading
			\hline
			$\mathrm{Al_2O_3}$ 	&	-0.43	&  92	&	-39.7	\\	
			fused silica		&	-0.03	&  10	&	-0.3	\\
			\hline
			\bottomrule
			\addstackgap[1pt]{\textbf{Total sensitivity}} &	&	& -69.0 \\
			\bottomrule
	\end{tabular}}
\end{table}

\begin{figure}[]
	\includegraphics[width=\columnwidth]{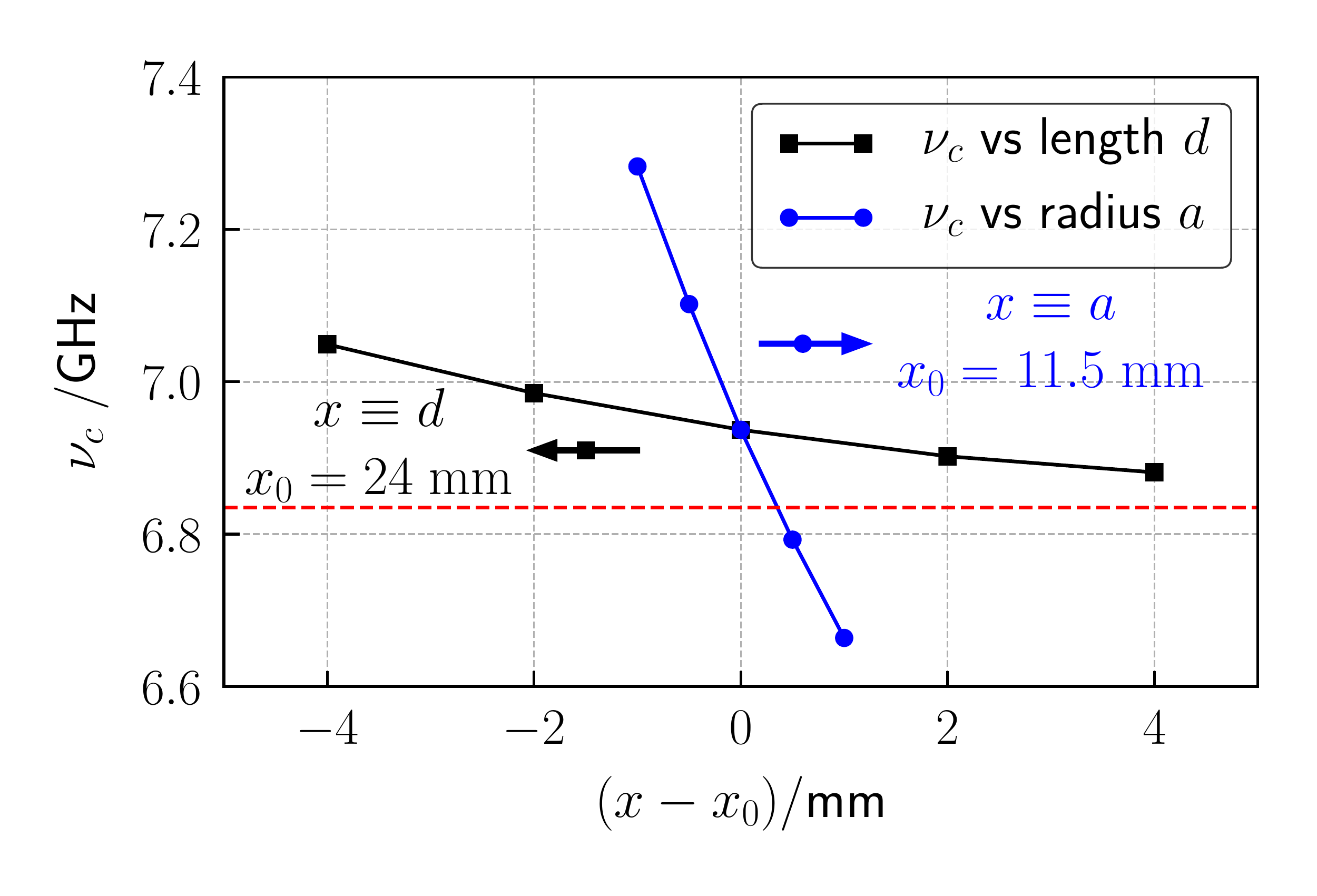}
	\caption{\label{fig:sweeps} $\mathrm{TE_{011}}$ resonance frequency as a function of cavity length and radius, as derived from the finite element analysis. The red dashed line corresponds to the atomic frequency ($\SI{6.8347}{\giga\hertz}$). }
\end{figure}

In \Cref{tab:temp_coeffs} the major contributions to the cavity frequency thermal sensitivity are expressed in relative terms. The sensitivity coefficients are taken from FEM calculations, that can provide a numerical evaluation of the mode frequency as a function of the various parameters. In \cref{fig:sweeps}, for example, the $\mathrm{TE_{011}}$-mode frequency is plotted as a function of the cavity length and radius. 

Close to the working points, we find that the main contribution from variations of the geometric parameters is due to the cavity radius: $\frac{1}{\nu_c}\frac{\Delta \nu_c}{\Delta T} = \SI[per-mode=symbol]{-23.9}{\ppm\per\kelvin}$. The sensitivity to the cavity thermal expansion is -1.17, close to the case of the empty cylindrical cavity cavity (equal to \num{-1}). Given the low thermal-expansion coefficient of alumina ($\alpha_{\mathrm{Al_2O_3}} = \SI[per-mode=symbol]{6.9}{\ppm\per\kelvin}$)~\cite{morgan-datasheet}), the geometric contribution from the loading material is instead small ($\simeq \SI[per-mode=symbol]{-2}{\ppm\per\kelvin}$).

The other major contribution to the temperature sensitivity comes from the variation of the dielectric constant of the alumina. From the manufacturer's datasheet~\cite{morgan-datasheet} we extract a thermo-dielectric coefficient $\beta_{Al_2O_3} = \SI[per-mode=symbol]{92}{\ppm\per\kelvin}$, while from the FEM analysis we obtain a linear dependence of the cavity frequency on $\epsilon$, with slope $\frac{\epsilon}{\nu_c}\frac{\Delta\nu_c}{\Delta\epsilon} = -0.43$. This value is close to the case of a fully loaded cavity cavity (-0.5). This is expected, since the electric field is mostly concentrated in the dielectric volume. The contribution from the linear expansion of the fused-silica cell and effects related to the Teflon spacers are instead negligible ($<\SI[per-mode=symbol]{0.1}{\ppm\per\kelvin}$).

Summing up all terms, the cavity resonance sensitivity is expected to be: $\Delta\nu_c/\Delta T = \SI[per-mode=symbol]{-472}{\kilo\hertz\per\kelvin}$. The total sensitivity to temperature is higher than the one of the unloaded cavity by a factor 3, but considering the lower filling factor (and for a typical cavity detuning of \SI{500}{\kilo\hertz} and loaded quality factor $Q_L=2800$) we estimate an expected clock fractional frequency sensitivity to temperature of $\simeq\SI[per-mode=symbol]{5e-12}{\per\kelvin}$ from cavity-pulling effect~\cite{Micalizio2009}. A temperature control at the level of \SI{0.5}{\milli\kelvin} is thus sufficient to reach state-of-the-art stability performance. We remind that given the small size of the cavity, such a level of temperature stabilization is not hard to achieve. 
\section{Experimental characterization}
\label{sec:exp}
\begin{figure}[t]
	\centering
	\includegraphics[width=0.7\columnwidth]{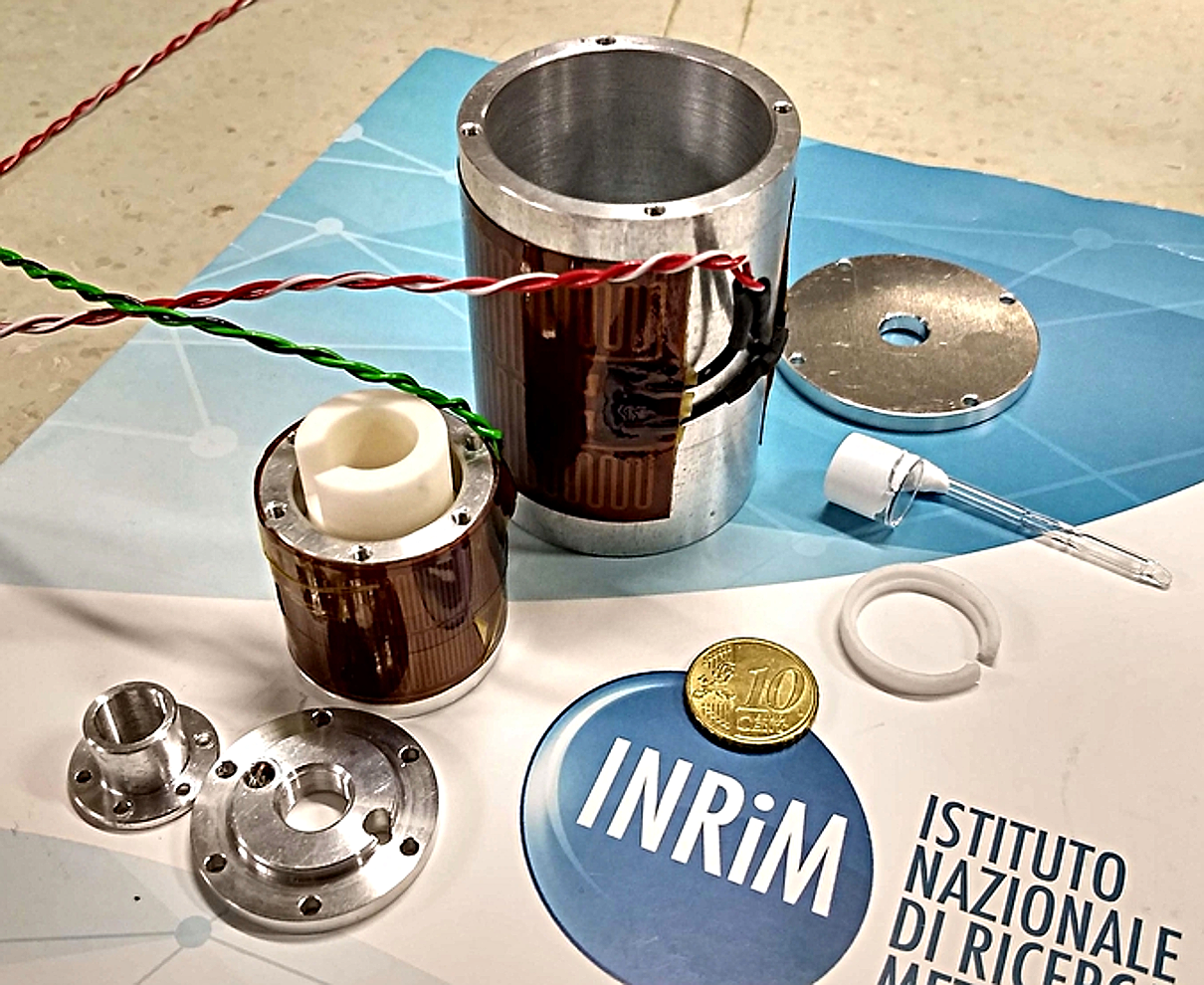}
	\caption{\label{fig:photo} Loaded cavity components prior to the assembly. In the picture it is also shown the first cylindrical thermal shield. As a dimensional reference, a 10 euro cent coin is shown.}
\end{figure}

The cavity-cell components are shown in \Cref{fig:photo} during the assembly phase. The cell used in the experimental validation is made of fused silica, with internal diameter and length both equal to \SI{1}{\centi\meter}. 

The loaded cavity resonance frequency has been experimentally measured by looking at the reflected power while sweeping the microwave frequency. 
The loaded cavity absolute frequency at the operational temperature of \SI{65}{\degreeCelsius} is found to be \SI{120}{\mega\hertz} lower than the simulated value. This is compatible with a dielectric constant for the alumina material closer to $9.7$, rather than the nominal value ($\epsilon=9.4$) which is provided by the manufacturer at $\SI{8.5}{\giga\hertz}$~\cite{morgan-datasheet}. A fine tuning has been achieved by adjusting the cavity length by fractions of \SI{1}{\milli\meter}. A measurement of the cavity resonance frequency at different temperatures lead to an experimental temperature coefficient of \SI[per-mode=symbol]{-473}{\kilo\hertz\per\kelvin} for the untuned cavity and of \SI[per-mode=symbol]{-461}{\kilo\hertz\per\kelvin} for the tuned cavity, in good agreement with the simulations. No significant degradation of the cavity intrinsic quality factor has been observed in the range \SIrange{25}{70}{\degreeCelsius}, due to excess deposition of metallic Rb. Indeed, $Q_i$ has a value of 3600 at room temperature, decreasing to 3200 around \SI{70}{\degreeCelsius}. 
\begin{figure}[]
	\includegraphics[width=\columnwidth]{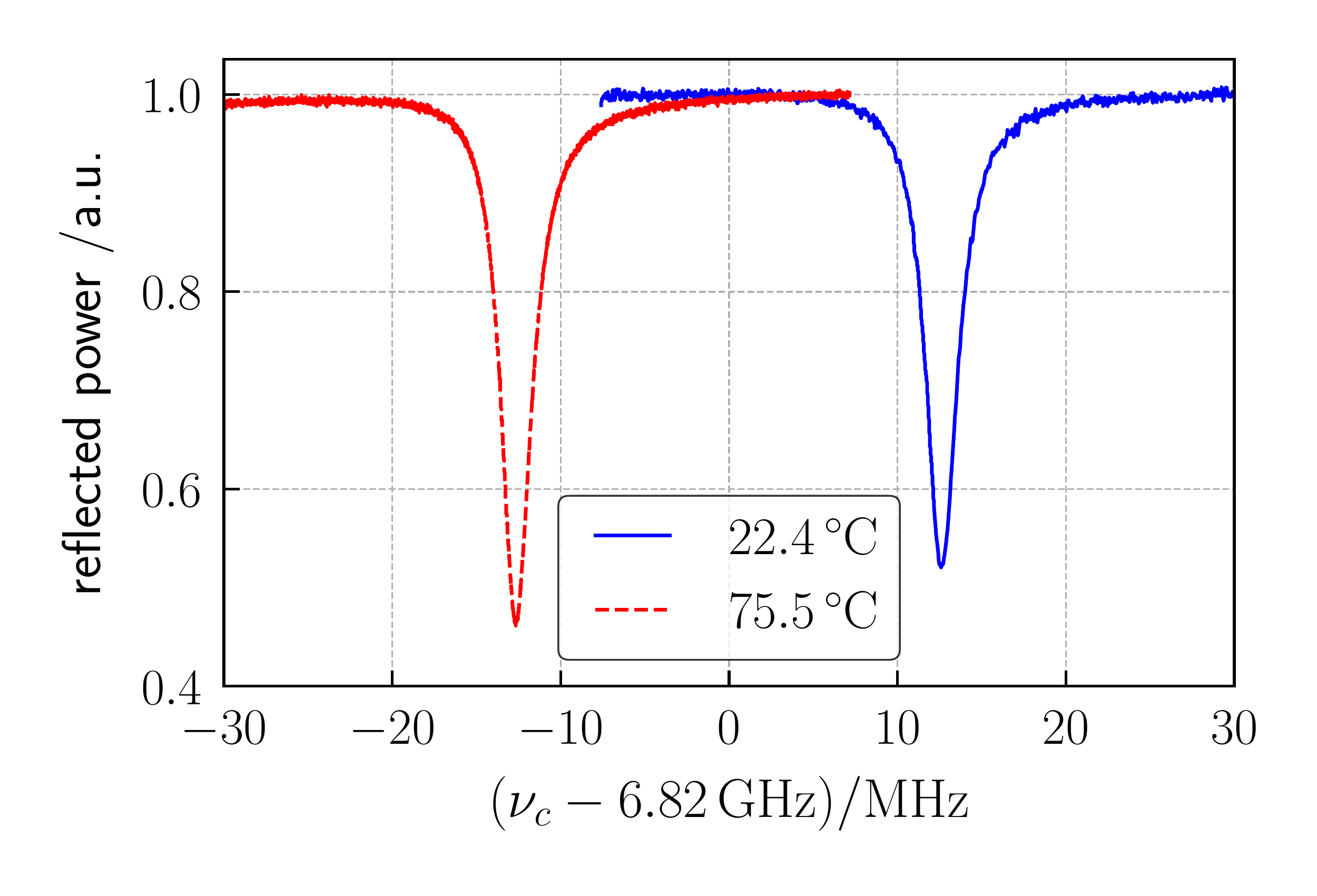}
	\caption{\label{fig:resonance} $\mathrm{TE_{011}}$ resonance frequency as a function of microwave detuning at two limit temperatures (ambient temperature and \SI{75.5}{\degreeCelsius}). The plot shows the reflected power from the cavity, detected with a microwave circulator and photodiode.}
\end{figure}
In \Cref{fig:resonance} the loaded-cavity resonance mode is shown for two limit temperatures. The measurements shown refer to the cavity resonance previous to the fine tuning. From this measurement we can also determine the cavity coupling parameter $\beta$~\cite{Godone2004}, which has a value of 0.2. 

\begin{figure}[]
	\includegraphics[width=\columnwidth]{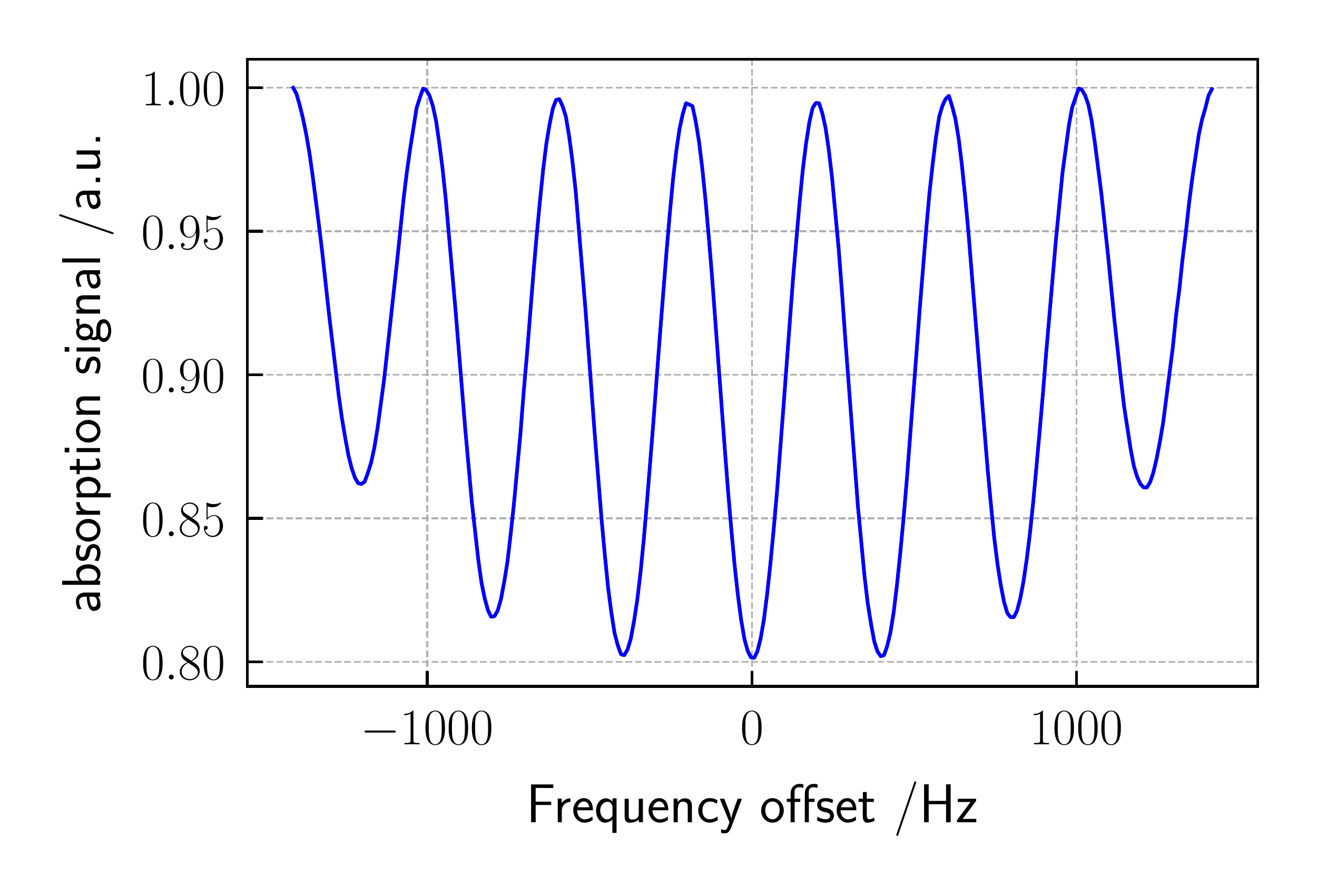}
	\caption{\label{fig:Ramsey} Ramsey scan of the atomic line (absorption signal as a function of the microwave detuning from the atomic frequency; each data point is the result of 3 averages). Cavity temperature \SI{64.5}{\degreeCelsius}, free-evolution time $T = \SI{2.0}{\milli\second}$, Rabi pulses length $t_1 = \SI{0.4}{\milli\second}$.}
\end{figure}

The proposed cavity-cell assembly performance has been tested with a POP clock scheme. The cavity has been operated at ambient pressure, integrated into an existing structure (same as in \cite{Calosso2007}) composed of 2 layers of thermal shielding and 3 layers of magnetic shielding. A static quantization magnetic field lifts the Zeeman degeneracy and isolates the clock transition. The optics package is the one described in \cite{Micalizio2012}, including a distribuited feedback laser (DFB) working on the $D_2$ line, frequency stabilized with an external reference cell through saturated absorption spectroscopy. The laser frequency is tuned to the minimum of the clock cell absorption profile for the $\ket{F=2}$ atomic ground-state. The pulsing is provided by an amplitude-modulated acousto-optic modulator operating in single-pass configuration.  The system is completed with a low-noise digital control and acquisition system~\cite{Calosso2017} and the microwave synthesis chain presented in \cite{Francois2015}. %Since only the short-term is under investigation at this stage, the thermal stability is not a critical aspect. 
In \Cref{fig:Ramsey} a scan of the Ramsey fringes, obtained with a laser absorption measurement as the microwave frequency is swept, is shown. The total cycle time is \SI{3.35}{\milli\second}, including pumping, clock interrogation and detection. The free-evolution time  $T=\SI{2}{\milli\second}$, while the Rabi pulses are $\SI{0.4}{\milli\second}$ long. The laser beam is collimated with a gaussian waist $2 w = \SI{6}{\milli\meter}$. The pumping power is \SI{4}{\milli\watt}, for a pulse duration of \SI{0.4}{\milli\second}. The detection pulse power and duration are \SI{200}{\micro\watt} and \SI{0.15}{\milli\second} respectively. With such timings and parameters, we obtained a fringe contrast of about \SI{20}{\percent}.

To characterize the atomic properties of the Rb sample in the clock cell, we measured the longitudinal relaxation time $T_1$, directly accessible with an optical detection of the population evolution, with the Franzen method~\cite{Franzen1959,Gharavipour2017}. By this means we get $T_1 = \SI{1.7(2)}{\milli\second}$ at \SI{64.5}{\degreeCelsius}. The transverse relaxation time $T_2$ is inferred by comparing envelope of the experimental Ramsey fringes to the one computed with the theory developed in \cite{Micalizio2013}. From this analysis, it turns out that $T_2$ is roughly $\SI{10}{\percent}$ lower than $T_1$, corresponding to $\gamma_2\simeq\SI{650}{\per\second}$.\\
The shift induced on the clock transition by the buffer gas is \SI[retain-explicit-plus]{+8555(5)}{\hertz}, consistent with a total buffer gas pressure of \SI{49(2)}{\torr}, assuming negligible error on the buffer gas composition~\cite{Vanier1982}. 

A finer tuning of the clock operating parameters (optical power, beam waist, cycle time, etc...) is needed, seeking for the ultimate stability performances.  However,  a stability below $\num{5e-13}$ at \SI{1}{\second} is compatible with the clock signal shown in \Cref{fig:Ramsey}, considering the major noise sources of the current setup (including laser RIN and frequency noise, detector noise, etc...). This estimate is confirmed by preliminary measurements and will be further characterized in future works.

\section{Conclusions}
\label{sec:conclusions}
A compact cavity-cell assembly, based on high-grade purity alumina as loading material has been designed, realized and characterized in terms of the main parameters of interest for microwave clock applications. The dielectric loading has lead to a reduction of a factor 10 of the inner cavity volume, compared to the traditional cylindrical cavity. The size reduction was achieved by maintaining the favorable field uniformity of the $\mathrm{TE_{011}}$ mode and high cavity quality factor. 

Despite the scaling of the clock cell dimension, introduced to push the miniaturization, we achieved high fringe contrast and high atomic-line quality factor by integrating the cavity in a POP clock setup. The obtained clock signal is compatible with a high-performing vapor-cell Rb clock, with short-term in the mid $10^{-13}\,\tau^{-1/2}$, as shown by preliminary characterization. The reduced size of the assembly facilitates thermal uniformity, with foreseen benefits in the medium-long term stability.    

The loaded-cavity approach thus adds to the existing design options for the realization of compact vapor cell clocks based on microwave interrogation. As shown in \cite{Wang2008}, more refined configurations can also be studied to make the design even more robust against environmental sensitivities. In this paper we reported a quality factor rather high for the POP with optical detection. If needed, this value can be tailored by using alumina with different percentages of impurities. 

The proposed loaded-cavity assembly paves the way for a strongly miniaturized Rb clock physics package with low mass and power consumption, particularly appealing for spaceborne applications.

\section*{Acknowledgments}
The authors thank Elio K. Bertacco for precious technical help and Marwan S.p.A. (Pisa, Italy) for the cell filling. We acknowledge valuable input on the alumina production process from J\"{o}rg-Uwe Wichert from Wesgo Ceramics (Erlangen, Germany) and Wesgo Ceramics GmbH for the alumina pieces procurement. We also thank the LED laboratory staff from Politecnico di Torino for providing the CST software and computational facility.

%\IEEEtriggeratref{17}

\bibliographystyle{ieeetr}
\bibliography{./biblio}

\end{document}